\documentclass[11pt]{article}
\usepackage{amsfonts}
\usepackage{amsmath, amssymb, amsthm}
\usepackage{fancyhdr}
\usepackage{graphicx}
\usepackage{dcolumn}

\begin{document}
\begin{center}
{\LARGE {\bf Reissner-Nordstrom black hole in dark energy background}}
\end{center}
\begin{center}
Ngangbam Ishwarchandra\dag1, \quad Ng. Ibohal\S \quad and K. Yugindro Singh\dag\\
\dag Department of Physics, Manipur University,\\ Imphal - 795003, Manipur, INDIA \\
\S Department of Mathematics, Manipur University,\\ Imphal - 795003, Manipur, INDIA\\
E-mail: \dag1 ngishwarchandra@manipuruniv.ac.in, \dag1 ngishwarchandra@gmail.com, \dag yugindro361@gmail.com\\ \S ngibohal@iucaa.ernet.in and \S ngibohal@manipuruniv.ac.in
\end{center}
\begin{abstract}
In this paper we propose a stationary solution of Einstein's field equations describing   Reissner-Nordstrom black hole in dark energy background. It is to be regarded as  the Reissner-Nordstrom black hole is embedded into the dark energy solution producing Reissner-Nordstrom-dark energy black  hole. We find that the space-time geometry of Reissner-Nordstrom-dark energy solution is Petrov type $D$ in the  classification of space-times. It is also shown that the embedded space-time possesses an energy-momentum tensor of the electromagnetic field interacting with the dark energy having negative pressure.  We find the energy-momentum tensor for dark energy violates the the strong energy condition due to the negative pressure, whereas that of the electromagnetic field obeys  the strong energy condition. It is shown that the time-like vector field for an observer in the Reissner-Nordstrom-dark energy space is expanding, accelerating, shearing and non-rotating.  We investigate the surface gravity of the horizons for the embedded dark energy black hole. The characteristic properties of relativistic dark energy based on the de Sitter solution is discussed in an appendix.\\\

\

\end{abstract}


\section{Introduction}
\setcounter{equation}{0}
\renewcommand{\theequation}{1.\arabic{equation}}
The standard general relativistic interpretation of dark energy is based on the cosmological constant [1], which has the simplest model for a fluid with the equation of state parameter $w=p/\rho=-1$, $\rho= - p=$ constant [2]. It turns out the cosmological constant to be the de Sitter solution with cosmological constant $\Lambda$ representing a relativistic dark energy with the non-perfect fluid energy-momentum tensor (A11) or elaborately (A9) having energy density $\rho=\Lambda/K$ and pressure $p=-\Lambda/K$ with $K=8\pi G/c^4$ (A8) (vide the appendix below). This suggests that a relativistic dark energy must have a line-element describing the gravitational field in the form of energy-momentum tensor having negative pressure with a minus sign in the equation of state parameter. It is also true that a vacuum space-time with $T_{ab}=0$ cannot have negative pressure to determine the equation of state parameter with minus sign. Hence the cosmological constant in {\it vacuum} Einstein's field equations cannot describe the negative pressure for the minus sign in the equation of state parameter. The criteria of minus sign in the equation of state is to indicate the matter distribution in the space-time  to be a dark energy, otherwise the plus sign in the equation of state is for the normal matter as discussed below. According to the de Sitter solution as relativistic dark energy (vide Appendix), there is another relativistic dark energy solution  admitting an energy-momentum tensor with the equation of state parameter $w=p/\rho=-1/2$, $\rho=4m/Kr$, $p=-2m/Kr$, where $m$ is a constant considered to be the mass of dark energy [3]. It is also to emphasize that the equation of state parameter $w=-1/2$ for the dark energy belongs to the range $-1<w<0$ focussed for the best fit with cosmological observations [4] and references there in. Here we shall refer this solution simply as dark energy solution without giving any extra prefix.

In general relativity the Schwarzschild solution is regarded as a black hole in an asymptotically flat space. Its generalization is the Reissner-Nordstrom black hole. The Reissner-Nordstrom-de Sitter solution is the charged extension of the Schwarzschild-de Sitter solution which is interpreted as a black hole in an asymptotically de Sitter space with the cosmological constant $\Lambda$ [5]. The Reissner-Nordstrom-de Sitter solution is also considered as an embedded black hole that the Reissner-Nordstrom solution is embedded into the de Sitter space to produce the Reissner-Nordstrom-de Sitter black hole [6]. Here we are looking for an exact solution to describe the Reissner-Nordstrom black hole in the dark energy with the parameter $w=p/\rho=-1/2$ as Reissner-Nordstrom-dark energy black hole. The embedded dark energy solution will be the generalization of Schwarzschild-dark energy black hole [7].

Here we consider the dark energy solution possessing a non-perfect fluid energy-momentum tensor having an equation of state parameter $\omega=p/\rho=-1/2$ with negative pressure derived in [3]. For deriving the Reissner-Nordstrom-dark energy solution we adopt the mass function expressed in a power series expansion of the radial coordinate [8] as
\begin{equation}
\hat{M}(u,r)= \sum_{n=-\infty}^{+\infty} q_n(u)\,r^n,
\end{equation}
where $q_n(u)$ are arbitrary functions of retarded time coordinate $u=t-r$. The mass function $\hat{M}(u,r)$ has a powerful role in generating new exact solutions of Einstein's field equations [9]. Wang and Wu [8] have utilized the mass function in deriving {\sl non-rotating} embedded Vaidya solution  into other spaces by choosing the function $q_n(u)$ corresponding to the index number $n$. Further utilizations of the mass function $\hat{M}(u,r)$ have been extended in rotating system and found the role of the number $n$ in generating rotating embedded solutions of the field equations [9]. Here we shall consider the cases of the index number $n$ as $n=0, -1, 2$. That the value $n = 0$ corresponds to the Schwarzschild solution, $n=-1$ for the charge term and $n = 2$ for the dark energy solution possessing the equation of state parameter $\omega=-1/2$ [3].
These values of $n$ will conveniently combine in order to obtain embedded charged Schwarzschild-dark energy black hole or Reissner-Nordstrom-dark energy black hole. The resulting solution will be an extension of the non-charged Schwarzschild-dark energy black hole discussed in [7], which is again a further extension of the work of [3] with the dark energy when $n=2$. Here we recall conveniently that the Reissner-Nordstrom-de Sitter black hole is the combination of two solutions corresponding to the index number $n=0, -1$ (Reissner-Nordstrom) and $n=3$ (de Sitter), different from the Reissner-Nordstrom-dark energy solution (to be discussed here) with $n=0, -1, 2$ in the power series expansion  of mass function $\hat{M}(u,r)$ (1.1).

The results of the paper are  summarized in the following theorems:
\newtheorem{theorem}{Theorem}
\begin{theorem}
The embedded Reissner-Nordstrom-dark energy solution admitting an energy-momentum tensor of electromagnetic field interacting with dark energy having an equation of state parameter $w=-1/2$, is a non-vacuum Petrov type  $D$ space-time.
\end{theorem}
\begin{theorem}
The equation of state parameter for dark energy has a minus sign, whereas the electromagnetic field (normal matter) has a plus sign showing the difference between the two matter fields.
\end{theorem}
\begin{theorem}
The time-like vector fields of the matter distribution in the Reissner-Nordstrom-dark energy space-time are expanding, accelerating and shearing with zero-twist.
\end{theorem}
\begin{theorem}
The energy-momentum tensor for dark energy possessing negative pressure violates the strong energy condition leading to a repulsive gravitational field.
\end{theorem}
\begin{theorem}
The energy-momentum tensor for the electromagnetic field (normal matter) having positive pressure satisfies the  strong energy condition providing an attractive gravitational field.
\end{theorem}
\begin{theorem}
The Reissner-Nordstrom-dark energy solution is expressible in Kerr-Schild ansatze on different backgrounds.
\end{theorem}

Theorem 1 shows the physical interpretation of the solutions that all  components of the Weyl tensors except $\psi_2$ of the space-time metric vanish indicating Petrov type D gravitational field. The energy-momentum tensors associated with the solution shows the interaction of electromagnetic field with the dark energy having the negative pressure and the energy equation of state parameters $w=-1/2$. Theorem 2 indicates the different characteristic properties of dark energy and the normal matter like electromagnetic field. Theorem 3 shows the physical interpretation of time-like vector field of the matter distribution in Reissner-Nordstrom-dark energy solution. Theorem 4 and 5 establish the characteristic features for the dark energy and the electromagnetic field possessing different gravitational fields. Theorem 6 strengthens the theorem stated in [10] that {\it every embedded black hole, stationary or non-stationary, is expressible
in Kerr-Schild ansatz}.
\vspace*{.15in}
\section{Reissner-Nordstrom-dark energy black hole}
\setcounter{equation}{0}
\renewcommand{\theequation}{2.\arabic{equation}}

In this section we shall show the derivation of an embedded Reissner-Nordstrom-dark energy solution to Einstein's field equations. This solution will describe charged Schwarzschild black hole in asymptotically dark energy background as the Reissner-Nordstrom-de Sitter black hole is regarded as a black hole in asymptotically de Sitter space [11]. For deriving an embedded Reissner-Nordstrom-dark energy solution, we choose the Wang-Wu function $q_{n}(u)$ in the expansion series of the mass function $\hat{M}(u,r)$  as
\begin{eqnarray}
\begin{array}{cc}
q_n(u)=&\left\{\begin{array}{ll}
M,&{\rm when }\;\;n=0\\
-e^2/2, &{\rm when }\;\;n=-1\\
m, &{\rm when}\;\;n=2\\
0, &{\rm when }\;\;n\neq 0,-1,2,
\end{array}\right.
\end{array}
\end{eqnarray}
where $M$ and $e$ are constants. Then, the  mass function takes the form
\begin{equation}
\hat{M}(u,r)=\sum_{n=-\infty}^{+\infty}
q_n(u)\,r^n =M+r^2m-\frac{e^2}{2r}.
\end{equation}
Then using this mass function in general canonical metric in Eddington-Finkestein coordinate system  $(u,r,\theta,\phi)$,
\begin{eqnarray*}
ds^2=\Big\{1-\frac{2\hat{M}(u,r)}{r}\Big\} du^2+2du\,dr-r^2d\Omega^2,
\end{eqnarray*}
with $d\Omega^2=d\theta^2+{\rm sin}^2\theta\,d\phi^2$, we find a line element
\begin{eqnarray}
ds^2&=&\Big[1-r^{-2}\Big\{2r(M+mr^2)-e^2\Big\}\Big]\,du^2+2du\,dr
-r^2d\Omega^2,
\end{eqnarray}
where $m$ is a constant regarded as the mass of the dark energy; and $M$ and $e$ denote the mass and the charge of Reissner-Nordstrom black hole. The line-element will reduce to that of Schwarzschild black hole when $m=e=0$ with singularity at $r=2M$, and also it will be that of dark energy when $M=e=0$ having singularity at $r=(2m)^{-1}$ [4].  The line-element (2.3) will have a singularity when $g_{uu}=0$, which has three roots $r=r_i$, $i=1, 2, 3$. The Reissner-Nordstrom solution has two roots $r=r_\pm$ for $g_{uu}=0$ when $m=0$. It is noted that the mass $m$ should not  be considered here to be zero for the existence of the dark energy.

The null tetrad components  for metric line element are obtained as follows
\begin{eqnarray}
&&\ell_a=\delta^1_a, \cr
&&n_a=\frac{1}{2}\Big[1-\frac{1}{r^2}\Big\{2r(M+mr^2)-e^2\Big\}\Big]\,\delta^1_a+ \delta^2_a, \cr
&&m_a=-{r\over\surd 2}\,\Big\{\delta^3_a +i\,{\rm
sin}\,\theta\,\delta^4_a\Big\}
\end{eqnarray}
with the normalization conditions $\ell_an^a= 1 = -m_a\bar{m}^a$ and other inner products being zero. The spin coefficients, Ricci scalars and Weyl scalars of the embedded Reissner-Nordstrom-dark energy black hole are found as follows:
\begin{eqnarray}
&&\kappa^*=\epsilon=\sigma=\nu=\lambda=\pi=\tau=0, \cr
&&  \rho^*=-{1\over r},\;\
\beta=-\alpha={1\over {2\surd 2r}}\,{\rm cot}\theta, \cr
&&\mu^*=-{1\over 2\,r}\Big\{1-{2M\over r}-2rm+{e^2\over r^2}\Big\}, \cr
&&\gamma={1\over 2\,r^2}\,\Big\{M-mr^2+\frac{e^2}{r}\Big\}, \\
&&\phi_{11}={1\over 4\,r^2}\Big\{2rm+e^2)\Big\},\cr
&&\Lambda^* = \frac{1}{2r}m, \cr
&&\psi_2={1\over 6\,r^4}(-rM+e^2). \nonumber
\end{eqnarray}
The above stationary space-time possesses an energy-momentum tensor describing the interaction of dark energy with the electromagnetic field as the source of gravitational field:
\begin{eqnarray}
T_{ab} &=& 2\,\rho\,\ell_{(a}\,n_{b)}
+2\,p\,m_{(a}\bar{m}_{b)},
\end{eqnarray}
where the quantities are found as
\begin{eqnarray}
&& \rho = {4\over Kr}m+\frac{e^2}{Kr^4}, \cr
&& p = -{2\over Kr}m+\frac{e^2}{Kr^4}.
\end{eqnarray}
These $\rho$ and $p$ are the density and pressure respectively for dark energy interacting with the electromagnetic field. Here $K$ denotes the universal constant $K=8\pi G/c^4$. The equation (2.7) indicates that the contribution of the gravitational field  to $T_{ab}$ is measured by dark energy  mass $m$ and the electric charge $e$ of Reissner-Nordstrom solution. The energy-momentum tensor (2.6) is calculated from Einstein's field equations $R_{ab} - (1/2)Rg_{ab} = -KT_{ab}$ of gravitational field for the space-time metric (2.3) with the relation $K\rho = 2\,\phi_{11} + 6\,\Lambda^*$ and $Kp = 2\,\phi_{11} - 6\,\Lambda^*$, where $\phi_{11}$ and $\Lambda^*$ are Ricci scalars [12] given in (2.5).

As in general relativity the physical properties of a space-time geometry  are  determined by the nature of the matter distribution in the space, it is convenient to express the energy-momentum tensor (2.6) in such a way that one must be able to understand it easily in order to study the physical properties of the embedded solution. Thus, the total energy-momentum tensor (EMT) for the solution (2.3) may, without loss of generality, be decomposed in the following form as:
\begin{eqnarray}
T_{ab} = T^{(\rm e)}_{ab}+T^{(\rm DE)}_{ab},
\end{eqnarray}
where the EMTs for the electromagnetic field $T^{(\rm e)}_{ab}$ and  the dark energy  $T^{(\rm DE)}_{ab}$ are  respectively given as:
\begin{eqnarray}
&&T^{(\rm e)}_{ab} =2\,\rho^{(\rm e)}\ell_{(a}\,n_{b)}
+2\,p^{(\rm e)}m_{(a}\bar{m}_{b)} \cr
&&T^{(\rm DE)}_{ab}=2\rho^{(\rm DE)}\,\ell_{(a}\,n_{b)} + 2p^{(\rm
DE)}\,m_{(a}\bar{m}_{b)}, \nonumber
\end{eqnarray}
where the coefficients are as
\begin{eqnarray}
&&\rho^{(\rm e)}= p^{(\rm e)}=\frac{1}{Kr^{4}}e^2, \\
&&\rho^{({\rm DE})}={4\over{Kr}}m,\quad
p^{(\rm DE)}=-{2\over Kr}m.
\end{eqnarray}
Thus, the equation of state parameters for the dark energy and the electromagnetic field are found  as
\begin{eqnarray}
&&\omega^{(\rm DE)}=\frac{p^{(\rm DE)}}{\rho^{(\rm DE)}}=-{1\over2} \\
&&\omega^{(\rm e)}=\frac{p^{(\rm e)}}{\rho^{(\rm e)}}= 1.
\end{eqnarray}
These two equations show that the equation of state parameter for dark energy has a minus sign in (2.11), whereas the  electromagnetic field (normal matter) has a plus sign (2.12) indicating the difference between dark energy and the normal matter. This follows the proof of Theorem 2.

The energy-momentum tensor (2.6) satisfies the energy conservation law [13] expressed in Newman-Penrose (NP) formalism [12]
\begin{eqnarray}
T^{ab}_{\;\;\;\,;b}=T^{({\rm e})ab}_{\;\;\;\:\;\:\;\;;b}+T^{({\rm DE})ab}_{\;\;\;\:\;\;\;\;\:\;\;\:;b}=0.
\end{eqnarray}
Here $T^{ab}$ itself satisfies the conservation law. On the other hand, we also find that both $T^{({\rm e})ab}$ and $T^{({\rm DE})ab}$ are separately satisfied the same (vide [7]). The above equation (2.13) shows the fact that the metric of the line element (2.3) describing embedded Reissner-Nordstrom-dark energy is a solution of Einstein's field equations.
The component of energy-momentum tensor (2.6) may also be written for future use as follows:
\begin{eqnarray}
T_{u}^{u}=T_{r}^{r}=\rho, \quad
T_{\theta}^{\theta}=T_{\phi}^{\phi}=-p.
\end{eqnarray}
We find the trace of the energy momentum tensor
$T_{ab}$ (2.6) as
\begin{equation}
T=2(\rho-p)=\frac{12}{Kr}m.
\end{equation}
Here we observe that $\rho - p$ must be always greater than zero for the existence of the dark energy in embedded Reissner-Nordstrom-dark energy solution (2.3) with $m \neq 0$, (if $\rho=p$ implies that $m$ will vanish). It is found that the charge $e$ does not appear in (2.15) showing the fact that the trace of the energy-momentum tensor for electromagnetic field always vanishes. The decomposition of energy-momentum tensor (2.8) indicates the interaction of electromagnetic field $T^{(\rm e)}_{ab}$ with the dark energy $T^{(\rm DE)}_{ab}$.

It is also convenient to write the energy-momentum tensor (2.6) in terms of  time-like $u_a$ $(u^au_a=1)$ and  space-like $v_a$ $(v^av_a=-1)$ vector fields as
\begin{eqnarray}
T_{ab}=(\rho+p)(u_{a}u_{b}-v_{a}v_{b})-pg_{ab}
\end{eqnarray}
where $u_a=(1/\sqrt{2})(\ell_a + n_a)$ and $v_a=(1/\sqrt{2})(\ell_a - n_a)$. This is certainly different from the energy-momentum tensor of the perfect fluid $T^{(\rm {pf})}_{ab} =
(\rho+p)u_a\,u_b-p\,g_{ab}$ with the trace $T^{(\rm {pf})}=\rho-3p$. We observe from the energy densities and the pressures given in (2.9) and (2.10) that the energy-momentum tensors for dark energy and electromagnetic field obey the weak energy and the dominant energy conditions given in [6].
However,  $T_{ab}^{(\rm DE)}$ violates the strong energy condition
\begin{eqnarray}
p^{\rm (DE)}\geq0,\quad \rho^{\rm (DE)}+p^{\rm (DE)}\geq0.
\end{eqnarray}
This violation of the strong energy condition is due to the negative pressure (2.10), and implies that the gravitational force of the dark energy is repulsive which may cause the acceleration of the model, like the cosmological constant leads to the acceleration of the expansion of the Universe. But, the energy-momentum tensor for electromagnetic field $T_{ab}^{(\rm e)}$ obeys the strong energy condition leading the attractive gravitational field
\begin{eqnarray}
p^{\rm (e)}\geq0,\quad \rho^{\rm (e)}+p^{\rm (e)}\geq0.
\end{eqnarray}
These complete the proofs of the Theorem 4 and 5 stated above for the two different matter fields possessing different gravitational forces.

The tetrad components of Weyl tensor $C^{a}_{bcd}$ of the embedded Reissner-Nordstrom-dark energy black hole (2.3) are found as
\begin{eqnarray}
&&\psi_0 =\psi_1 =\psi_3 =\psi_4 =0, \cr
&&\psi_2 =\frac{1}{r^4}\Big\{-rM+ e^2\Big\}.
\end{eqnarray}
The non-vanishing Weyl scalar $\psi_2$ indicates that the space-time of the embedded solution  (2.3) is Petrov type $D$ in the classification of space-times. The mass $m$ of the dark energy does not appear in  (2.19) which shows the intrinsic property  of the conformally flatness of the dark energy, even embedded into the Reissner-Nordstrom black hole. From (2.3), (2.11) and (2.19) it follows the proof of the Theorem 1 stated in the introduction above.

The curvature invariant for the Reissner-Nordstrom-dark energy black hole (2.3) is found as
\begin{eqnarray}
R_{abcd}R^{abcd}&=&\frac{48}{r^8}(-Mr+e^2)^2+\frac{8}{r^8}(mr^3+e^2)^2+\frac{24}{r^2}m^2.
\end{eqnarray}
This invariant diverges only at the origin $r=0$. This indicates that the origin $r=0$ is a physical singularity. This shows that the singularity of the Reissner-Nordstrom-dark energy solution (2.3) is caused due to the coordinate system, such like in Schwarzschild solution.

\subsection{Kerr-Schild ansatze of Reissner-Nordstrom\\-dark energy metric}
Here we shall clarify the nature of the embedded solution in the form of Kerr-Schild ansatze in different backgrounds that every embedded solution is expressible in Kerr-Schild ansatz [10]. The Reissner-Nordstrom-dark energy metric can be expressed in
Kerr-Schild ansatz on the dark energy  background
\begin{equation}
g_{ab}^{\rm (RNDE)}=g_{ab}^{\rm (DE)} +2Q(r)\ell_a\ell_b
\end{equation}
where $Q(r) =-Mr^{-1}+e^2r^{-2}/2$. Here,
$g_{ab}^{(\rm DE)}$ is the dark energy metric and $\ell_a$ is  geodesic, shear free, expanding and zero twist null vector for
both $g_{ab}^{(\rm DE)}$ as well as $g_{ab}^{(\rm RNDE)}$. The above
Kerr-Schild form can also be recast on the Reissner-Nordstrom background
as
\begin{equation}
g_{ab}^{(\rm RNDE)}=g_{ab}^{(\rm RN)} +2Q(r)\ell_a\ell_b
\end{equation}
where $Q(r) = - mr$.
 These two
Kerr-Schild forms (2.21) and (2.22) show the fact that the Reissner-Nordstrom-dark energy space-time (2.3) with the mass $m$ of the dark energy is a solution of Einstein's field equations. In view of the Kerr-Schild ansatze (2.21) and (2.22) we have two different nomenclatures -- Reissner-Nordstrom-dark energy and dark energy-Reissner-Nordstrom. That ``Reissner-Nordstrom is embedded into the dark energy space to obtain Reissner-Nordstrom-dark energy black hole $g_{ab}^{\rm (RNDE)}$ (2.21)'' or ``the dark energy space-time is embedded into the Reissner-Nordstrom black hole to produce the dark energy-Reissner-Nordstrom black hole $g_{ab}^{\rm (DERN}$ (2.22) -- both  nomenclatures possess the same geometrical structure. This is the important characteristic property of embedded solutions that it is hard physically to distinguish which space of the two has started first to embed into another.
However, in the context of combination of exact solutions, it is to emphasize the fact that the two metrics $g_{ab}^{\rm (RN)}$ for Reissner-Nordstrom solution and $g_{ab}^{(\rm DE)}$ for dark energy cannot be added in order to obtain $g_{ab}^{\rm (RNDE)}$ as
\begin{equation}
g_{ab}^{\rm (RNDE)}\neq \frac{1}{2}\Big\{g_{ab}^{\rm (RN)} + g_{ab}^{(\rm DE)}\Big\}.
\end{equation}
It is the fact that in general relativity two physically known solutions cannot be added to derive a new embedded solution. From (2.21) and (2.22) we come to the conclusion of the proof of theorem 6 above.

\subsection{Raychaudhuri equation of Reissner-Nordstrom-dark energy \\solution}
We shall study the nature of the time-like vector $u_a=(1/\sqrt{2})(\ell_a + n_a)$ appeared in the energy-momentum tensor (2.16) for the Reissner-Nordstrom-dark energy solution (2.3). In fact it describes the physical properties of the matter whether the matter is expanding ($\Theta=u^a_{\:\,;a} \neq 0$), accelerating ($\dot{u}_a=u_{a;b}u^b\neq 0$) shearing $\sigma_{ab}\neq0$ or non-rotating ($w_{ab}=0$). We shall investigate the rate of expansion from the Raychaudhuri equation, such that we can understand how the negative pressure of the dark energy affects the expansion rate of the model.
We write the explicit form of the time-like vector as
\begin{eqnarray}
u_a&=&\frac{1}{2\sqrt{2}}\Big[3-\frac{2}{r}\Big(M+r^2m-\frac{e^2}{2r}\Big)\Big]\delta^1_a +\frac{1}{\sqrt{2}}\delta^2_a.
\end{eqnarray}
This expression of $u_{a;b}$ is convenient to obtain the expansion scalar $\Theta=u^a_{\:\,;a}$ and acceleration vector $\dot{u}_a=u_{a;b}u^b$ and the shear tensor $\sigma_{ab}=\sigma_{(ab)}$ as follows
\label{eq:whole}
\begin{eqnarray}
\Theta &=& \frac{1}{\sqrt{2}\,r^2}\Big(r+M+3r^2m\Big), \\ \label{subeq:1}
\dot{u}_a&=&\frac{1}{\sqrt{2}r^2}\Big\{M-mr^2-\frac{e^2}{r}\Big\}v_a \\ \label{subeq:2}
\sigma_{ab}&=&\frac{1}{3\surd 2 r^2}\Big(r+4M-\frac{3e^2}{r}\Big)(v_av_b-m_{(a}\bar{m}_{b)}),  \label{subeq:3}
\end{eqnarray}
which is orthogonal to $u^a$ ({\it i.e.,} $\sigma_{ab}u^b=0$).
We find that for the solution (2.3), the vorticity tensor $w_{ab}$ is vanished.
It is observed that the mass $m$ of the dark energy does not explicitly involve in the expression  of $\sigma_{ab}$ but its involvement can be seen in the expression of $u_a$. However, the mass $m$ directly involves in the expression of the expansion (2.25) as well as the acceleration (2.26). We find from (2.26) that the Reissner-Nordstrom-dark energy solution discussed here follows the non-geodesic path of the time-like vector ($u_{a;b}u^b\neq 0$). This establishes the key result of the expansion of the solution with acceleration. The vanishing of the vorticity tensor $w_{ab}=0$ can be interpreted physically  as saying that the Reissner-Nordstrom-dark energy solution is non-rotating ({\it twist-free}) as mentioned earlier.

Now we shall investigate the consequences of the Raychaudhuri equation for the solution (2.3). The Raychaudhuri equation states
\begin{equation}
\dot{\Theta}=\dot{u}^a_{\:\,;a}+2(w^2-\sigma^2)-\frac{1}{3}\theta^2 +R_{ab}u^a u^b
\end{equation}
where $\dot{\Theta}=\Theta_{;a}u^{a}$; the shear and vorticity magnitudes are $2\sigma^2=\sigma_{ab}\sigma^{ab}$ and $2w=w_{ab}w^{ab}$ respectively; $R_{ab}$ is the Ricci tensor of the line element (2.3). Then the Raychaudhuri equation for our  twist-free time-like vector
is found as follows
\begin{eqnarray}
\dot{\Theta}&=&-\frac{1}{4r^2}(1+2mr)-\frac{M}{r^4}(r+M+mr^2)+\frac{e^2}{4r^4}\Big(1+\frac{14M}{r}\Big).
\end{eqnarray}
From this we observe that when the black hole mass $M$ and the charge $e$ vanish, the rate of change of the expansion $\dot{\Theta}$ will be that of the dark energy solution. The first term of (2.29) is for the dark energy solution, the first and second terms are for the Schwarzschild-dark energy solution [6]; all three terms correspond for the Reissner-Nordstrom-dark energy solution showing the interaction of dark energy and the electromagnetic field. Hence, the equations (2.25), (2.26) and (2.27) establish the proof of Theorem 3 that the time-like vector fields of the matter distribution in the Reissner-Nordstrom-dark energy space-time are expanding, accelerating and shearing with zero-twist.
\subsection{Surface Gravity of Reissner-Nordstrom-dark energy black hole}
In this subsection we shall discuss the surface gravity of Reissner-Nordstrom-dark energy black hole on the horizon. The equation $\Delta=0$ of the embedded Reissner-Nordstrom-dark energy black hole (2.3) will provide the horizons
\begin{equation}
\Delta\equiv 1-r^{-2}\Big\{2r(M+mr^2)-e^2\Big\}=0.
\end{equation}
This equation has three roots $r_1$, $r_2$ and $r_3$ that $r_1$ is real, other $r_2$ is complex and $r_3$ the conjugate of $r_2$. They are explicitly found as
\begin{eqnarray}
r_1&=&\frac{1}{6m} + H - G \\
r_2&=&\frac{1}{6m} - \frac{(1+i \sqrt{3})}{2} H +\frac{(1-i \sqrt{3})}{2}G \cr
r_3&=&\frac{1}{6m} - \frac{(1-i \sqrt{3})}{2} H +\frac{(1+i \sqrt{3})}{2}G, \nonumber
\end{eqnarray}
where
\begin{eqnarray}
H&=&\frac{1}{6\times 2^{1/3}m\,g}(-1+12mM), \quad
G=\frac{1}{6\times 2^{1/3}m}g, \cr
g&=&\Big\{h+\sqrt{4(-1+12mM)^3+h^2}\,\Big\}^{1/3}, \cr
h&=&-2\{1+54m^2e^2-18mM\}.
\end{eqnarray}
These roots satisfy the relation
\begin{eqnarray}
(r-r_1)(r-r_2)(r-r_3)=-\frac{r^2}{6m}\Delta,
\end{eqnarray}
where $\Delta\equiv 1-r^{-2}\{2r(M+mr^2)-e^2\}$.
The real root $r_1$ describes the horizon of the Reissner-Nordstrom-dark energy black hole, as the other complex roots $r_2$ and $r_3$ may have less physical interpretation.

The surface gravity $\kappa$ of the Schwarzschild-dark energy black hole on the horizons $r=r_1$ can be obtained as follows
\begin{eqnarray}
\kappa=\frac{1}{2}\Delta'={1\over r^2}\Big\{M-mr^2-\frac{e^2}{r}\Big\}\Big|_{r=r_1}.
\end{eqnarray}
It is observed that the charge $e$ of Reissner-Nordstrom black hole affects the surface gravity (2.34) of the embedded Reissner-Nordstrom-dark energy black hole. This indicates the interaction of dark energy distribution with the electromagnetic field. The surface gravity has an important feature that it determines the Hawking temperature $T=\kappa/2\pi$ of the embedded black hole on the horizons $r=r_1$.

\section{Conclusion}
In this paper we proposed an exact solution of Einstein's field equations describing the Reissner-Nordstrom black hole embedded into the dark energy space having negative pressure as Reissner-Nordstrom-dark energy black hole. This embedded solution is the straightforward generalization of Schwarzschild-dark energy solution [6]. Here we have followed the method of generating embedded solutions of Wang and Wu [8] by considering the power index $n$ as $n=0, -1$ and $2$ in the derivation of the solution. Then we calculate all the NP quantities for the line element and find that the embedded space-time possesses an energy-momentum tensor of the electromagnetic field interacting with the dark energy having negative pressure. We have shown the difference between the dark energy and the normal matter (like electromagnetic field) that dark energy has the equation of state parameter with minus sign, whereas the normal matter has the parameter with plus sign.
The energy-momentum tensor of the dark energy distribution in the embedded space-time (2.3) violates the strong energy condition leading to a repulsive gravitational force, whereas that of the electromagnetic field satisfies the strong energy condition producing attractive gravitation field. The metric tensor of Reissner-Nordstrom-dark energy solution is able to express in Kerr-Schild ansatze on different backgrounds (2.21)  and (2.22) establishing the fact that the Reissner-Nordstrom-dark energy space-time (2.3) with the mass $m$ of the dark energy is a solution of Einstein's field equations. We investigate the physical properties of the solution and present them in the form of theorems with proofs.

It is observed that the polynomial equation $\Delta=0$ (2.30) for the Reissner-Nordstrom-dark energy solution has three roots $r=r_i$, $(i=1, 2, 3)$. When the charge $e$ becomes zero, the polynomial will be $\Delta\equiv 1-2r^{-1}(M+mr^2)=0$ for the Schwarzschild-dark energy solution  having two roots $r_\pm =(1/4m)\{1\pm\sqrt{1-16mM}\,\}$ [7]. Further if we set the black hole mass $M$ to zero, it will be that of the dark energy solution possessing one root $r=(2m)^{-1}$ [3]. Here according to Bousso [14], stars are as distance as billions of light years, so $r>10^{60}$ and stars are as old as billions of years, $t>10^{60}$. In this scale, the size of the mass of the dark energy at the horizon $r=(2m)^{-1}$ may become
\begin{eqnarray}
m=\frac{1}{2}r^{-1}<\frac{1}{2}\times10^{-60},
\end{eqnarray}
which is bigger than the size of the de Sitter cosmological constant $|\Lambda|\leq3r_{\Lambda}^{-2}\leq 3\times10^{-120}$ with the cosmological horizon $r_\Lambda=\sqrt{3/\Lambda}$. As  mentioned earlier it is also to emphasize that the equation of state parameter $w=-1/2$ for the dark energy belongs to the range $-1<w<0$ focussed for the best fit with cosmological observations [4] and references there in.

We find that the time-like vector of the observer is expanding $\Theta \neq 0$ (2.25), accelerating $\dot{u}_a\neq 0$ (2.26) as well as shearing $\sigma_{ab} \neq 0$ (2.27), but non-rotating $w_{ab}=0$. The equation (2.26) for $\dot{u}_a=u_{a;b}u^b \neq 0$ means that the stationary observer of the solution does not follow the time-like geodesic path  $u_{a;b}u^b = 0$. We also find that the energy-momentum tensor for the dark energy solution violates the strong energy condition (2.17). The violation of strong energy condition is due to the negative pressure of the dark energy in the space-time geometry, and is not an assumption in order to obtain the other dark energy models [15].

The black  hole mass $M$ does not involve in the Ricci scalar $\phi_{11}$ showing the vacuum character of the Schwarzschild solution and the charge $e$ in the $\Lambda^*\equiv(1/24)g^{ab}R_{ab}$ indicating the traceless of energy-momentum tensor for the electromagnetic field. Also the dark energy mass $m$ does not involve in the Weyl scalar $\psi_2$ confirming the conformal flat character of the dark energy solution [7]. The Ricci scalar $\phi_{11}$ shows the interaction of electromagnetic field and dark energy as both the charge $e$ and the dark energy mass $m$ are appeared in one expression. The decomposition of the energy-momentum tensor (2.8) shows the possibility of the interaction of  dark energy with the electromagnetic   field. The interaction of dark energy with other matter fields is in agreement with the suggestion made by Sahni and Starobinsky [16]. It is also to mention that when the black hole mass $M=0$, the energy-momentum tensor with the decomposition (2.8) will remain unaffected i.e., in that situation the space-time (2.3) will describe a {\it charged dark energy} solution having the same properties given in (2.11), (2.12), (2.13), (2.15), (2.17) and (2.18).

The decomposition (2.8) of energy-momentum tensor (2.6) indicates the interaction of electromagnetic field with the dark energy. This is one of the remarkable properties of the Reissner-Nordstrom-dark energy that two different matters  of distinct physical properties are present in one energy-momentum tensor (2.6) as the source of gravitational field. It is also seen that the trace of $T_{ab}$ is $T=2(\rho-p)=2(\rho^{(\rm DE)}-p^{(\rm DE)})$ which is different from that of perfect fluid $T^{(\rm {pf})}=\rho-3p$. The energy-momentum tensor for the dark energy with negative pressure violates the strong energy condition while that for the electromagnetic field with positive pressure obeys the condition showing the difference between the dark energy and the normal matter (electromagnetic field). In fact this embedded solution possessing a non-perfect fluid energy-momentum tensor may be an example of space-times which are enable to explain how the dark energy is different from the normal matter. Another example of embedded space-time possessing a non-perfect fluid energy-momentum tensor is the Reissner-Nordstrom-de Sitter solution as the de Sitter cosmological constant $\Lambda$ is regarded as a candle of dark energy having an equation of state parameter $\omega^{(\rm dS)}={p^{(\rm dS)}}/{\rho^{(\rm dS)}}=-1$. In this paper we have seen different characters of dark energy and the normal matter possessing different gravitational fields in a single energy-momentum tensor of the space-time geometry.

\section*{Acknowledgments}

One of the authors (Ibohal) is thankful the Inter University Centre for Astronomy
and Astrophysics (IUCAA), Pune for hospitalities during his visit.

\section*{Appendix A: de Sitter solution as relativistic dark energy with cosmological constant}
\setcounter{equation}{0}
\renewcommand{\theequation}{A\arabic{equation}}
Here we shall show, by using the expansion series of the mass function [8], the derivation of de Sitter solution of cosmological constant $\Lambda$ which is interpreted as the simplest example of dark energy [4], [15], [16] and discuss the properties of the solution possessing the equation of state parameter $w=-1$ with negative pressure. So that we shall compare the properties of the dark energy solution, discussed here in this paper, possessing the equation of state $w=-1/2$ which is in the range $-1<w<0$ focussed for the best fit with cosmological observations [4] and references there in.

We consider a line-element of a general canonical metric in Eddington-Finkestein coordinate systems $\{u, r, \theta, \phi\}$
\begin{equation}
ds^2=\Big\{1-\frac{2}{r}M(u,r)\Big\} du^2+2du\,dr-r^2\{d\theta^2+{\rm sin}^2\theta\,d\phi^2\},
\end{equation}
where  $M(u,r)$ is referred to as the mass function and related to the gravitational fields within a given range of radius $r$. Here $u=t-r$ is the retarded time coordinate.

Using the metric tensor of the line-element in the Einstein's field equations $R_{ab} - (1/2)Rg_{ab} = -KT_{ab}$, we obtain an energy-momentum tensor describing the matter field distribution in the gravitational field as
\begin{eqnarray}
T_{ab}=\mu\ell_a\ell_b+2\,\rho\,\ell_{(a}\,n_{b)}
+2\,p\,m_{(a}\bar{m}_{b)},
\end{eqnarray}
where the quantities are found as
\begin{eqnarray}
&&\mu=-\frac{2}{Kr^2}M(u,r)_{,u}, \quad
\rho = \frac{2}{Kr^2}M(u,r)_{,r}, \cr\cr
&& p = -\,\frac{1}{Kr}M(u,r)_{,rr}
\end{eqnarray}
with the universal constant $K=8\pi G/c^4$. The energy-momentum tensor (2.7) is calculated from Einstein's equations $R_{ab} - (1/2)Rg_{ab} = -KT_{ab}$ of gravitational field for the space-time metric (2.3) with the relation $K\,\mu^* = 2\,\phi_{22}$, $K\rho = 2\,\phi_{11} + 6\,\Lambda^*$ and $Kp = 2\,\phi_{11} - 6\,\Lambda^*$, where $\phi_{11}$, $\phi_{22}$ and $\Lambda^*\equiv(1/24)\,R_{ab}\,g^{ab}$ are NP Ricci scalars [12]. Here $\ell_a$, $n_a$, $m_a$ and $\bar{m}_a$ are  the complex null tetrad vectors given as
\begin{eqnarray}
&&\ell_a=\delta^1_a, \cr
&&n_a=\frac{1}{2}\Big\{1-\frac{2}{r}M(u,r)\Big\}\,\delta^1_a+ \delta^2_a, \cr
&&m_a=-{r\over\surd 2}\,\Big\{\delta^3_a +i\,{\rm
sin}\,\theta\,\delta^4_a\Big\},
\end{eqnarray}
$\bar{m}_a$ being the complex conjugate of $m_a$.
For obtaining the de Sitter solution with cosmological constant $\Lambda$,
we choose the Wang-Wu function for the index number $n=3$ in (1.10) above.
\begin{eqnarray}
\begin{array}{cc}
q_n(u)=&\left\{\begin{array}{ll}
\Lambda/6, &{\rm when}\;\;n=3\\
0, &{\rm when }\;\;n\neq 3,
\end{array}\right.
\end{array}
\end{eqnarray}
such that the mass function $M(u,r)$ takes the form
\begin{equation}
M(u,r)=\frac{1}{6}r^3\Lambda.
\end{equation}
Then using this in the line element (A1) we obtain the de Sitter solution with the cosmological constant $\Lambda$ in Eddington-Finkestein coordinate systems $\{u, r, \theta, \phi\}$ as
\begin{eqnarray}
ds^2&=&\Big\{1-\frac{1}{3}r^2\Lambda\Big\}\,du^2 +2du\,dr
-r^2(d\theta^2+{\rm sin}^2\theta\,d\phi^2).
\end{eqnarray}
From (A3) and (A6) we have the quantities  as follows
\begin{eqnarray} \mu =0  \quad
\rho = \frac{\Lambda}{K}, \quad p =
-\frac{\Lambda}{K}.
\end{eqnarray}
where $\rho$  and $p$ are the energy density  and pressure  of the
de Sitter solution with cosmological constant $\Lambda$. Then the energy-momentum tensor (A2) becomes
\begin{eqnarray}
T_{ab}=2\,\rho\,\ell_{(a}\,n_{b)}
+2\,p\,m_{(a}\bar{m}_{b)},
\end{eqnarray}
having the trace as $T=2(\rho-p)$. This is the  energy-momentum tensor for the de Sitter solution with the cosmological constant $\Lambda$. From (A8) we obtain the equation of state parameter as the ratio of the pressure to the density which takes the form
\begin{eqnarray}
w=\frac{p}{\rho}=-1.
\end{eqnarray}
Replacing the energy density $\rho$ and pressure $p$ of (A8) in (A9) and using the metric tensor $g_{ab} =2\,\ell_{(a}n_{b)} -2\,m_{(a}\bar{m}_{b)}$ [12], we can write the  energy-momentum tensor in the most familiar form for the de Sitter metric
\begin{eqnarray}
KT_{ab}=\Lambda\,g_{ab}.
\end{eqnarray}
This is possible only because of the non-perfect fluid energy-momentum tensor (A9); otherwise it may not be possible to write it at all. Then the Einstein's field equations $R_{ab} - (1/2)Rg_{ab} = -KT_{ab}$ can be written in the familiar form with cosmological constant $\Lambda$ as
\begin{eqnarray}
R_{ab} - (1/2)Rg_{ab}+\Lambda\,g_{ab}=0.
\end{eqnarray}
These field equations do not mean the vanishing energy-momentum tensor as long as the relativistic dark energy de Sitter solution associated with $T_{ab}$ (A11) is concerned having negative pressure with the equation of state parameter $w=p/\rho=-1$ in the space-time. However, the field equations (A12) may be interpreted as the Einstein's  field equations for a vacuum space-time with a cosmological constant $\Lambda$. In this respect Durrer and Maartens [1] have noted that ``a vacuum energy and a classical cosmological constant cannot be distinguished by observation.'' It is also emphasized that the  energy-momentum tensor (A9) for de Sitter solution is different from that of a perfect fluid,  $T^{(\rm {pf})}_{ab} =
(\rho+p)u_au_b-p\,g_{ab}$ with a unit time-like vector $u_a$ and
the trace $T^{(\rm {pf})}=\rho-3p$. The de Sitter space-time (A7) is {\it conformally flat} with $C^{a}_{\;\;bcd}=0$,  {\it i.e.} all the tetrad components of Weyl tensor are  vanished $\psi_L=0$, $L=0, 1, 2, 3, 4$.

Here we also mention that the strong energy condition for the energy-momentum tensor of the form (A9) or (2.6) above for a non-perfect fluid has  the following condition [7], [13]
\begin{eqnarray}
p\geq0,\quad \rho+p\geq0.
\end{eqnarray}
This energy condition is violated for the de Sitter model due to the negative pressure (A8) leading to a repulsive gravitational field for a dark energy solution [3].

It is also to mention for more information  about the de Sitter cosmological constant that the
Einstein's field equations for the de Sitter model with the cosmological function $\Lambda(u)$ take the following form [13]
\begin{eqnarray}
R_{ab} - \frac{1}{2}\,R\,g_{ab}  + \Lambda(u)g_{ab} =
-T_{ab}^{(\rm NS)},
\end{eqnarray}
where the non-stationary evolution part $T_{ab}^{(\rm NS)}$ of the energy-momentum
tensor is given by
\begin{eqnarray*}
T_{ab}^{(\rm NS)}=-\frac{1}{3}r\Lambda(u)_{,u}\ell_a\ell_b,
\end{eqnarray*}
which has zero-trace $T^{(\rm NS)}=0$ and is vanished when the cosmological function $\Lambda(u)$ becomes constant. This shows that the introduction of variable $\Lambda(u)$ in the field equations needs to care the effect in the change of energy-momentum tensor (A2). However, the equation of state parameter $w$ has the same form as in (A10) with the negative pressure $p = -\Lambda(u)/K$ and the density $\rho = \Lambda(u)/K$.

From the above we observe that the de Sitter solution with cosmological constant $\Lambda$, which possesses the energy-momentum tensor (A9) with negative pressure (A8) and having the equation of state parameter $w=p/\rho=-1$ (A10), is one of the relativistic dark energy solutions. The energy-momentum tensor (A9) does not describe the perfect fluid and violates the strong energy condition (A13) due to the negative pressure. Hence, the dark energy solution with the equation of state parameter $w^{(\rm DE)}=p^{(\rm DE)}/\rho^{(\rm DE)}=-1/2$ discussed here in this paper is interpreted as another example of relativistic dark energy solutions. It is also to state for further reference that the non-perfect fluid energy-momentum tensor (2.6) or (A9) is a general form of $T_{ab}$ for stationary solutions including dark energy [3], Reissner-Nordstron-dark energy (2.6), Reissner-Nordstron-de Sitter, rotating de Sitter, rotating monopole, Kerr-Newman-de Sitter [9].

\end{document}